\documentclass[11pt,twoside]{article}

\usepackage{asp2006}
\usepackage{epsf}
\usepackage{lscape}
\usepackage{graphicx}

\begin{document}

\title{Star formation histories of resolved stellar populations: in and beyond the Local Group}   
\author{Monica Tosi}   
\affil{INAF - Osservatorio Astronomico di Bologna, Via Ranzani 1, I-40127 Bologna, Italy}  

\begin{abstract} 
The exploitation of the power and the spatial resolution of HST and new
generation ground-based telescopes allows to measure with good precision 
the individual stars of galaxies in and beyond the Local Group. This leads 
to very successful studies of the star formation
histories of galaxies of different morphological types. Our current knowledge 
of the
star formation history of galaxies within 10-20 Mpc, as derived from the
colour-magnitude diagrams of their resolved stellar populations, is reviewed
here.
\end{abstract}


\section{Introduction}   
\begin{figure}[!ht]
\begin{center}
\includegraphics[width = 280pt, height = 200pt]{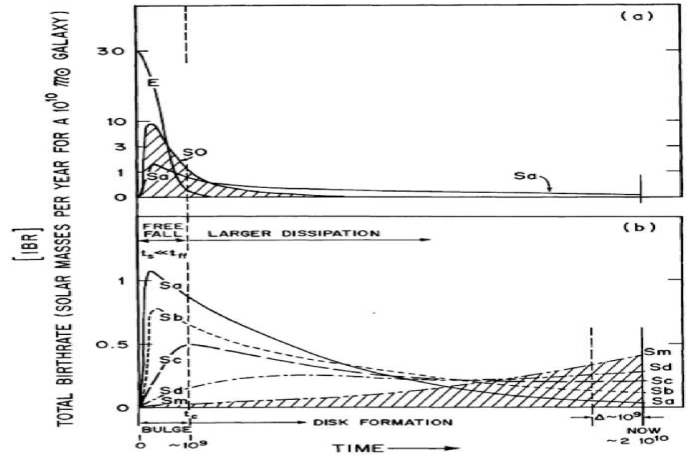}
\end{center}
\caption{Star formation rate as a function of time in galaxies of different
morphological type, from \citet{S86}.
}
\label{sandage}
\end{figure}

The research field of star formation histories (SFHs) in resolved galaxies is 
one of those which have advanced the most in the last 15 -- 20 years.
In 1986, Alan Sandage presented his enlightening view of how the 
star formation rate (SFR) varies with time in galaxies of different
morphological type (Fig.\ref{sandage}).  From a collection of  
heterogeneous indicators (chemical abundances, line emissions, colors, etc.), 
he understood that the star formation (SF) activity
is very strong and concentrated at the earliest epochs in the earliest-type
galaxies and has lower and lower intensity, but longer and longer timescale
going toward later morphological types.  Magellanic irregulars may even
have, on the long term, an average SFR increasing with time. 

The best tool to trace the SFH of a galactic region is  the 
colour-magnitude diagram (CMD) of its resolved stellar populations, 
where the signatures of all their evolutionary properties (age, metallicity, 
initial mass function - IMF) remain imprinted. When \citet{S86} presented his 
perspective view of the SFH, the available telescopes and instruments had only
recently attained photometric performances adequate to resolve individual 
stars in nearby galaxies. A few people were already working 
at the development of the procedures that are now known as the synthetic CMD 
method to derive the SFH, whose results are summarized in the next sections. 
We were pioneers, with few tools at our disposal and we should all be grateful 
to Cesare Chiosi for his contribution to the advance of this field and for 
the fundamental tools his group has made available, such as the Padova sets of 
homogeneous stellar evolution tracks covering all masses from 0.6 to 120
M$_{\odot}$ and several initial metallicities. 

\section{The synthetic CMD method}

The method allows to derive the SFH parameters (epoch and duration of the SF 
episodes and of the quiescent phases, SFRs, IMF) of a stellar system from its 
observational CMDs. It is based on the comparison of the observational CMDs  
with theoretical CMDs created via Monte~Carlo extractions on stellar 
evolution tracks, or isochrones, for a variety of SFHs, IMFs, 
and age-metallicity relations \citep[for details, see e.g.][]{T91,G98,ag04}. 
The synthetic CMDs take into account the number of stars of the observational 
CMD (or portions of it), photometric errors, incompleteness and blending 
factors. Hence, a combination of assumed parameters is acceptable only 
if the resulting synthetic CMD reproduces all
the features of the observational one: morphology, colours, luminosity 
functions, number of stars in specific evolutionary phases.
Since this is a statistical approach, the method does not provide unique 
solutions for the SFH of the examined region, but strongly reduces
the range of possible scenarios.

The observational CMDs obtainable 15 years ago from the existing instrumentation 
nowadays don't look very appealing; yet the first applications of the synthetic 
CMD
method immediately proved its power and its capability of providing interesting
new perspectives \citep{F89,T91,Be92,Ga96,T96}. We found that the SFH differs
significantly from one region to another, even in tiny galaxies such as the
Local Group (LG) dwarf irregular (dIrr) WLM
\citep{F89}, and that the SF activity in late-type dwarfs usually occurs in 
long episodes of moderate intensity, separated by short quiescent phases, rather
than in short episodes of strong intensity, separated by long quiescent
intervals, contrary to what
people used to think of the active dwarfs. In other words a 
{\it gasping} rather than a {\it bursting} regime \citep{M95}. 

Then HST became available and boosted a tremendous interest in the SFH research
field. HST provided
such tight and deep CMDs that many more groups became interested in the
synthetic CMD method and developped their own procedures to derive the SFHs.
Already in 2001, when Carme Gallart organized the famous 
experiment in Coimbra (Portugal) to compare with each other the results 
from the synthetic CMD methods of different groups, ten groups participated to 
the experiment \citep[and references therein]{SG02}. The importance of that
initiative is that it allowed to find out strengths and weaknesses of the
different procedures and showed that, within the uncertainties, most procedures
provided consistent results. The experiment consisted in deriving the SFH of a
region of the LMC bar from \citet{Sme02} HST/WFPC2 data, and we all found for 
that region a fairly continuous SF activity, although with variable strength. 
The inferred SFH of the LMC
bar turned out to be quite different from that of the LMC clusters
\citep[e.g.][]{PT98}, as shown in the bottom-right panel of Fig.\ref{sfall},
where the empty histograms refer to the clusters and the filled ones to our
\citep{To02} results on the Coimbra experiment.

\section{Resolved stellar populations in and beyond the Local Group}

Nowadays, it is possible to resolve individual  faint/old stars  
in galaxies of the LG and its immediate vicinities, and we can thus 
infer their SFHs over 
long lookback times (up to the Hubble time), with an average time resolution
around 10\% of the lookback time. The high spatial resolution of 
new generation imagers (the HST cameras {\it in primis}) also allows us to 
spatially resolve the SFH not only in the Milky Way satellites but in all nearby
galaxies. For instance \citet{DP98} and \citet{DP02} have  resolved
and measured the SF activity over the last 0.5 Gyr in the various sub-regions 
of the dIrrs Gr8 and Sextans A, close to the borders of the LG.  
The resulting space/time distribution of the SF, with lightening and fading of 
adjacent cells, is
intriguingly reminiscent of the predictions of the stochastic self-propagating
SF theory proposed by \citet{SSG79} almost 30 years ago.
 
The HST/ACS provides spectacularly deep and spatially resolved images, such
as those of the star forming region NGC346 in the SMC, where \citet{No06} 
and \citet{S07} have been able to measure 85000 stars, 
from very old to very young ones, including for the first time 
pre-main-sequence objects of mass from 3 to 0.6 M$_{\odot}$. 
The impressive CMD obtained from HST/ACS images of M31 by \citet{Br03} has
opened the way to accurate studies of the resolved stellar populations of other
spiral galaxies (see also A. Ferguson and T. Smecker-Hane, this volume, for M31
and M33).

In galaxies beyond the LG, distance makes crowding more severe, and even 
HST cannot resolve stars as faint as the  main-sequence turn-off (MS-TO) of old
populations. The higher the distance, the worse the crowding conditions, and the
shorter the lookback time reachable even with the deepest, highest resolution
photometry.  Depending on distance and intrinsic crowding, the reachable 
lookback time in galaxies more than 1 Mpc away ranges from several 
Gyrs (in the best cases, when the red giant branch or even the horizontal
branch are clearly identified), to several hundreds Myr (when asymptotic giant
branch stars are recognized), to a few tens Myr (when only the brightest
supergiants are resolved). 

\begin{figure}[!ht]
\begin{center}
\includegraphics[scale=0.3]{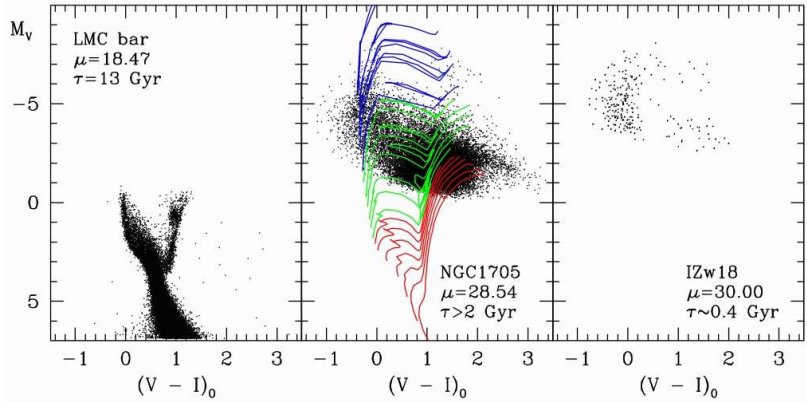}
\end{center}
\caption{Effect of distance on the resolution of individual stars and on the
corresponding lookback time. The panels display the CMDs of 
systems observed with the HST/WFPC2 and analysed with the same techniques, 
but at different distances. 
The adopted instrinsic distance modulus $\mu$ is indicated in each panel,
as well as the lookback time $\tau$. In the central panel, the Padova
evolutionary tracks for metallicity Z=0.004 \citep{F94} are also shown to help
in recognizing the various evolutionary phases. .
}
\label{dist}
\end{figure}

The effect of distance on the capability of resolving individual stars and
therefore on the reachable lookback time is shown in Fig.\ref{dist}, 
where the CMDs obtained from HST/WFPC2 photometries of three late-type 
galaxies are shown: the LMC
bar \citep{Sme02}, with a distance of 50 kpc and a CMD reaching a few mags
below the old MS-TO, i.e. a lookback time of 13 Gyr; NGC1705 \citep{To01}, 
with distance 5.1 Mpc and a CMD
reaching a few mags below the tip of the red giant branch (RGB), i.e. a lookback
time of more than 2 Gyr; and IZw18 \citep{Al99}, with a very uncertain 
distance (here assumed at the lowest limit of 10 Mpc) where we reached only 
the asymptotic giant branch (AGB), i.e. stars at least 0.4 Gyr old. 

\begin{figure}[!ht]
\centerline{
\scalebox{1.0}{%
\includegraphics[width = 120pt, height = 138pt]{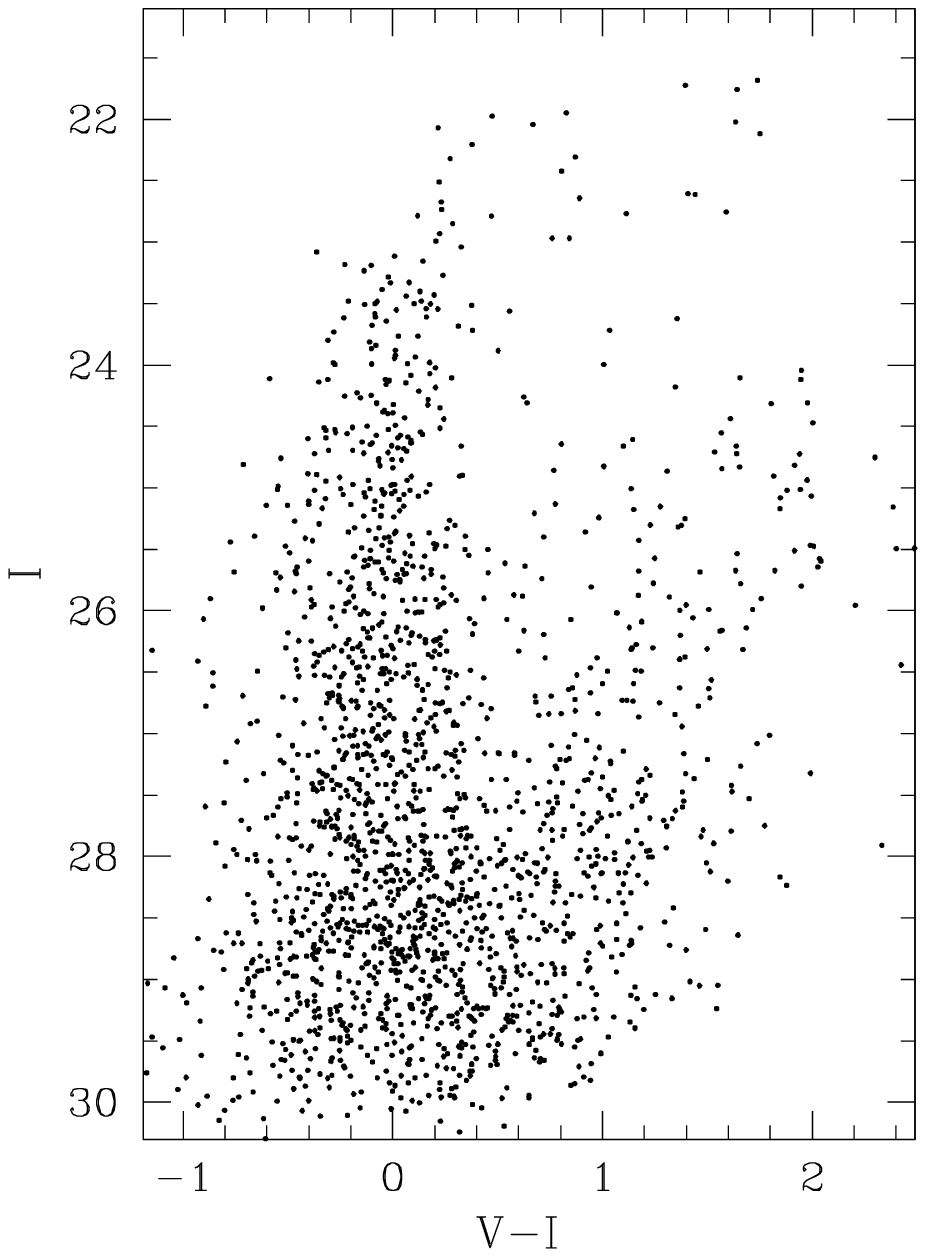}}
\scalebox{1.0}{%
\includegraphics[width = 240pt, height = 135pt]{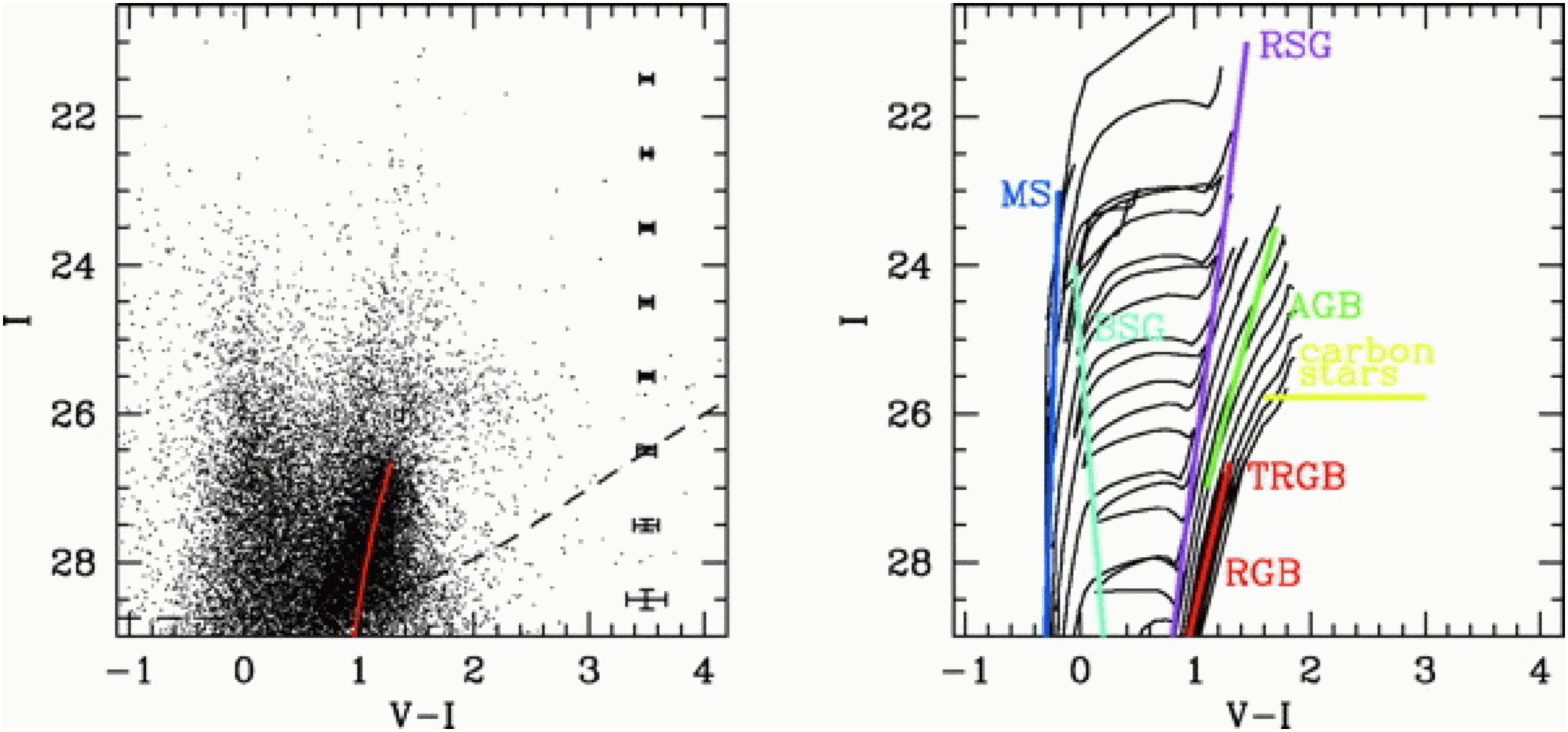}
}
}
\caption{CMDs of the most metal-poor BCDs resolved with HST/ACS: IZw18 and 
SBS1415+437 (left and central panel, respectively). 
The right-hand panel shows \citet{Gir02}
Z=0.001 isochrones, shifted at the distance of SBS1415+437. See text for details
and references.
}
\label{sbs}
\end{figure}

The problem is that not all kinds of galaxies are present in the LG, with 
the unfortunate circumstance that missing are systems representative of the
most extreme
morphological types, ellipticals and blue compact dwarfs (BCDs). We must then 
deal with the higher uncertainties and study the SFH of galaxies also outside 
the LG. The existing instrumentation currently allows to resolve AGB stars and,
often, RGB stars in galaxies within 15 - 20 Mpc. For example, 
Fig.\ref{sbs} shows the CMDs from HST/ACS photometry of two of the three most
metal-poor BCDs ever discovered: IZw18 \citep{To06} and SBS1415+437 
\citep{Al05}, both more distant than 10 Mpc. 

In the case of SBS1415+437 (central panel in Fig.\ref{sbs}) 
a prominent RGB is clearly measured, 
demonstrating that the galaxy contains stars older than about 2 Gyr. The 
luminosity of the RGB tip implies a distance of 13.6 Mpc \citep{Al05}.
By comparing the CMD with the Padova isochrones of 
appropriate metallicity (Z=0.001), shifted at that distance, it is apparent 
that SBS1415+437 also contains intermediate age AGB stars (including the 
thermally pulsing phase and carbon stars) and young massive stars. 
This leads to the important conclusion that, although very blue, currently 
active, metal poor, and gas rich, this system has not started only recently to
form stars but has managed to sustain a significant star formation 
activity over more than 2 Gyr.

The case of IZw18 \citep[left-hand panel in Fig.\ref{sbs}, from][]{To06} 
is more complicated,
because crowding is more severe and the distance may be higher. Yet we do 
see in its CMD an increase in the stellar density along the red branch below
I $\simeq$ 27.5, which is most likely due to the appearance of the RGB, specially
once we consider that photometric incompleteness is increasingly affecting the
faintest magnitudes \citep[see also][]{Mom05}. New HST/ACS time series 
observations,
suitable for the derivation of Cepheids light-curves (Aloisi et al. in
preparation) are providing an independent estimate of the galaxy distance and
confirming the detection of the RGB. These results show that, contrary to 
previous claims \citep[e.g.][]{IT04}, even the most metal-poor star forming 
galaxy ever discovered cannot be considered a genuinely young system and is 
instead active since at least a couple of Gyr. Moreover, its extremely low
metallicity (Z=0.0004 for the youngest stars, presumably less for 
the older ones) significantly reduces the age-metallicity degeneracy on the
RGB, and its colour extension therefore traces a significant age spread,
suggesting that the onset of the SF activity in IZw18 has occurred several Gyrs
ago.

\section{Star formation histories}

The SFHs of LG galaxies derived from their resolved stellar populations were
extensively described by \citet{Gr98} and \citet{Ma98}. Since then, many other
studies have been performed, applying the synthetic CMD method to a large
fraction of the LG dwarf galaxies, both of late and early-type.
Very interesting results  on the SFH
are expected from the ACS images of M31 (A. Ferguson, this volume), M33 (T.
Smecker-Hane, this volume) and a number of selected dwarfs of various types
(e.g. C. Gallart, this volume). 
Many other groups have worked on the SFHs of nearby dwarf and giant galaxies,
but it is unfortunately impossible to cite them all here.  

\begin{figure}[!ht]
\centerline{
\scalebox{1.0}{%
\includegraphics[width = 180pt, height = 220pt]{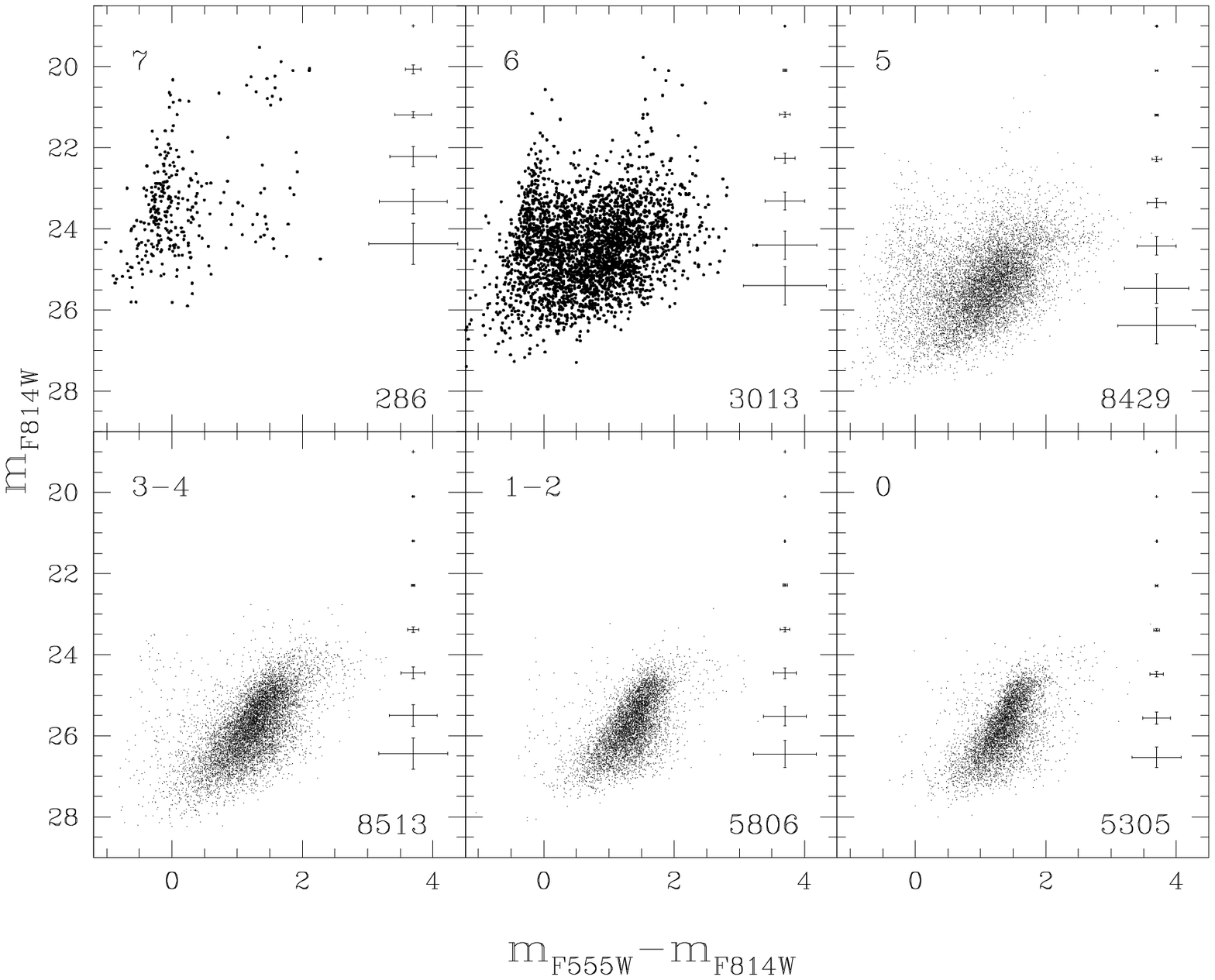}}
\scalebox{1.0}{%
\includegraphics[width = 180pt, height = 215pt]{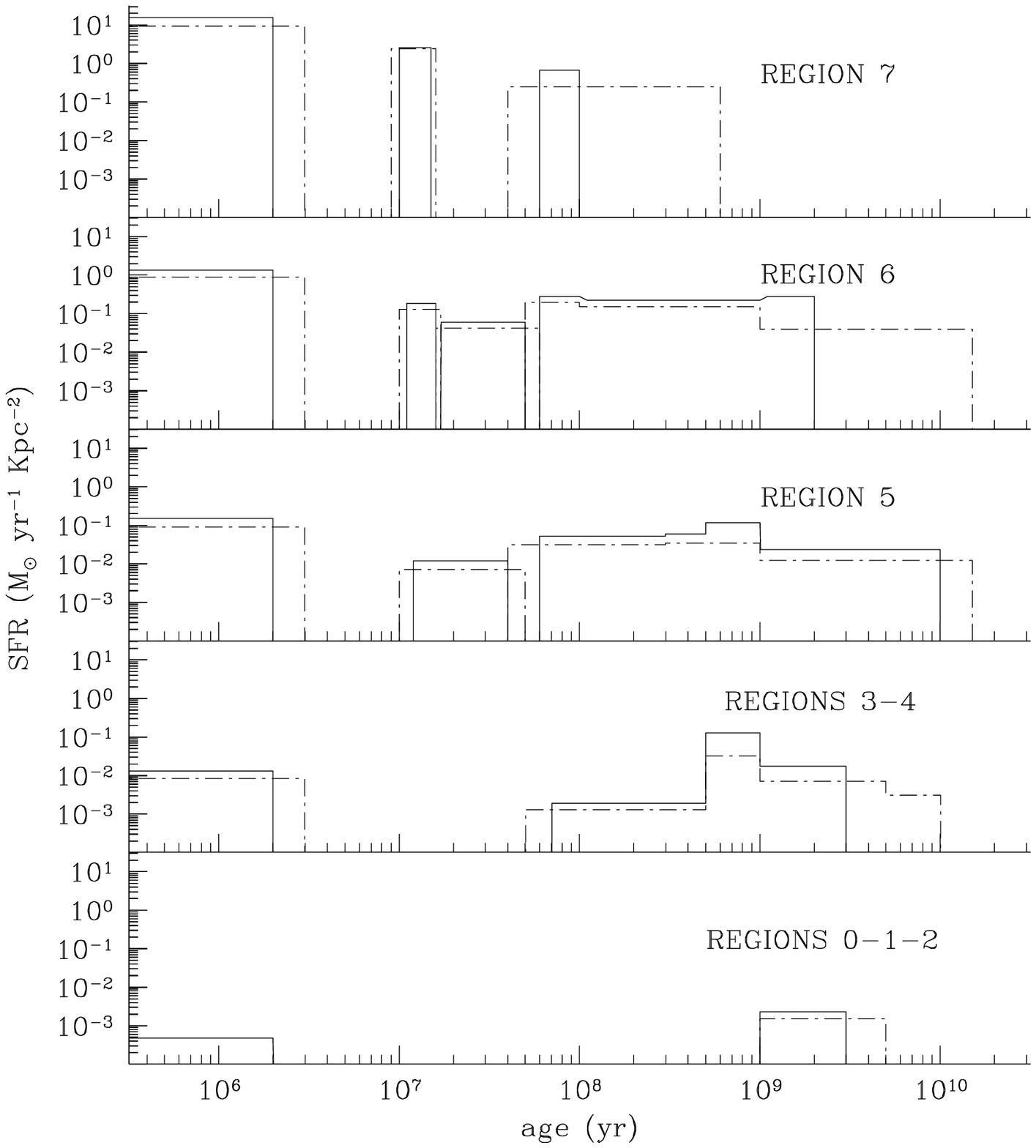}
}
}
\caption{The CMDs of concentric regions of the BCD NGC1705 \citep{To01} are 
plotted in the left-hand panels, where the photometric errors and the number of
resolved stars are indicated. Regions from 7 through 0 span from the galactic
center to the outskirts. The right-hand panels show the
corresponding SFH (SFR density vs age), and the different line-types refer to
alternative scenarios \citep[see][for details]{An03}.
}
\label{n1705}
\end{figure}

Not many groups have embarked in the more challenging application of
the synthetic CMD method beyond 1 Mpc, 
and most of them have concentrated their efforts on
the understanding of late-type dwarfs \citep[e.g.][]{Va96,L98,sl01,c03,Va05}, 
with notable exceptions, such as the study of the recent SFH in the elliptical 
galaxy NGC5128 \citep{Re04}.   
Our group has studied some of the most active dwarfs outside the LG: the dIrr
NGC1569 and the BCDs NGC1705, IZw18 and SBS1415+437. 

Fig.\ref{n1705} shows the case of NGC1705, a very active BCD at 5.1 Mpc, with 
evidence of galactic winds. The most central region is very crowded and only
young and intermediate-age stars can be resolved, but elsewhere RGB stars are
easily resolved and become the most prominent CMD feature in the outer regions.
We can then derive the SFH of most regions back to several Gyr ago, as shown in
the right-hand panel of Fig.\ref{n1705}. It is immediately apparent that in
this BCD we are not seeing a {\it bursting} regime, with short episodes of
SF activity separated by long periods of quiescence. It is also interesting to
notice that the strong SF episode of 15 Myr ago coincides with the epoch of
both the formation of the central super star cluster and the onset of the 
galactic wind. The wind is probably the cause of the sudden halt of the SF
activity everywhere in NGC1705. The gas, however, must have been quick in
cooling and falling down, since only 6-7 Myr later another, stronger, SF burst
appears to occur in all the galaxy.

\begin{figure}[!ht]
\begin{center}
\includegraphics[scale=0.6]{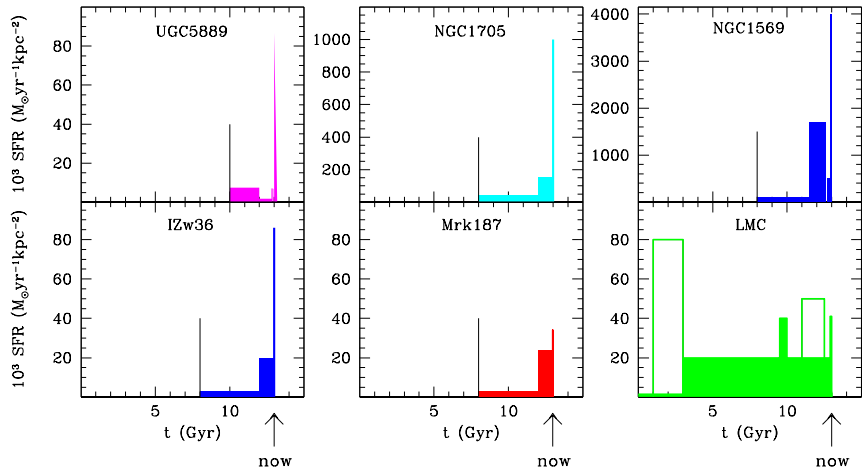}
\end{center}
\caption{SFHs in late-type galaxies derived with the synthetic CMD method. The
thin vertical line in each panel indicates the reached lookback time. See text
for references.
}
\label{sfall}
\end{figure}

Fig.\ref{sfall} summarizes the SFHs derived for six late-type galaxies, all, but
the LMC, more than 1 Mpc away: the LMC bar \citep{To02}, the dIrr NGC1569
\citep{G98,An05}, the BCDs NGC1705 \citep{An03}, IZw36 \citep{sl01} and
Mrk187 \citep{sl00}, and the LSB UGC5889 \citep{Va05}. The most important result
is that all of them were already active at the reached lookback time. All of
them present a recent SF activity, which is what let people discover them in
spite of the distance. None of them exhibits long quiescent phases within the
reached lookback time. On the other hand, the SFR differs significantly from one
galaxy to the other, independently of the attributed morphological type. It is
true that the least active object is the LSB galaxy, as expected, but the most
active one is by far the dIrr NGC1569. Of the BCDs, two have SFRs comparable to
that of the LMC bar, and the third doesn't reach the SFR level of NGC1569.

The statistics  is still poor, but we can already draw some general results from
all the SFHs derived so far:
\par\noindent
{\bf 1.} Evidence of long interruptions in the SF activity is found only in 
early-type galaxies;
\par\noindent
{\bf 2.} No galaxy currently at its first SF episode has been found yet;
\par\noindent
{\bf 3.} No frequent evidence of strong SF bursts is found in late-type dwarfs;
\par\noindent
{\bf 4.} There is no significant difference in the SFH of dIrrs and BCDs, except
for the current SFR.

\acknowledgements 
I am grateful to Cesare Chiosi for his invaluable contribution to this research
field: he has obtained very interesting results, he has 
been one of the first few who understood the importance of complete and
homogeneous sets of stellar evolution models and he has made them available. 
Most of the results described here have been obtained thanks to nice 
collaborations with A. Aloisi, L. Angeretti, F. Annibali, L. Greggio,
C. Leitherer, A. Nota, and E. Sabbi. 
Part of these researches was funded through INAF-PRIN-2005.



\end{document}